\magnification=\magstep 1
\overfullrule=0pt
\hfuzz=16pt
\voffset=0.0 true in
\vsize=8.8 true in
\baselineskip 12pt
\parskip 6pt
\hoffset=0.1 true in
\hsize=6.3 true in
\footline={\hfil -- {\folio} -- \hfil}
 
\centerline{\bf Time Scales of Phonon Induced Decoherence of Semiconductor Spin Qubits}

\vskip 0.32in

\centerline{ D. Mozyrsky$^1$, Sh. Kogan$^2$, and G. P. Berman$^1$}

\vskip 0.24in

\noindent{$^1$ T-13 and CNLS, Los Alamos National Laboratory,
Los Alamos, NM 87545}

\noindent{$^2$ MST-11,Los Alamos National Laboratory,
Los Alamos, NM 87545}

\vskip 0.32in

\centerline{\bf Abstract}

Decoherence of a shallow donor electron spin in silicon caused by electron-lattice interaction is studied. We find that there are two time scales associated with the evolution of the electron spin density matrix: the fast but incomplete decay due to the interaction with non-resonant phonons followed by slow relaxation resulting from spin flips accompanied by resonant phonon emission. We estimate both time scales as well as the magnitude of the initial drop of coherence for P donor in Si and argue that the approach used is suitable for evaluation of phonon induced decoherence for a general class of localized spin states in semiconductors.       

\bigskip


\vfil\eject

\bigskip

Rapid development of quantum information science has revived interest in study of electron and
nuclear spin polarization, specifically in condensed matter physics. Potentially powerful applications
of electronic and nuclear spins in quantum computing [1] and spintronics [2] have recently been proposed. Among these
quantum information devices the ones utilizing spins in semiconductors seem to be the most promising candidates for practical implementation. Indeed the existing semiconductor technology may someday allow for scalability of the elementary quantum logic gates into complex integrated circuits needed for information processing. 

The problem of fabrication of the semiconductor quantum processing devices opens up a number of experimental and theoretical challenges. These include control problems, such as manipulations by interactions of individual spins with external field and with each other [3] (in the present case the $1/2$-spins represent bits of information and will be conventionally referred as qubits), the initialization and readout of the quantum register [4], as well as precise positioning of the qubits (within several $\AA$) in a semiconductor host - the issue that requires significant advances made in contemporary lithographic technologies. 

Another requirement for implementation of semiconductor devices for quantum information processing is a long time memory of the state of a quantum register, i.e. small decoherence of the spin qubits due to the interaction with environment. Several theoretical works that study decoherence in systems contemplated for quantum information processing have been made recently [5]. There is an outstanding set of experiments on spin relaxation processes in GaAs systems [2,6]. It was discovered that at certain doping concentration the transverse relaxation time $T_2$ of the electron spins injected into conduction band by optical pumping techniques increases by several orders of magnitude and exceeds $100 \, {\rm ns}$ under certain conditions. This remarkable phenomenon suggests possibilities of its use in the new generation of electronic nano-devices that utilize polarized spins as information carriers and can become the building blocks of the future electronics.
       
In the present work we focus on study of decoherence of {\it localized} electronic spins in semiconductors. The localized electronic or nuclear spins have been proposed to play role as qubits in several quantum computing architectures [7]. It has been long known [8-10] that in certain semiconductors (such as Si or Ge) the spins of electrons bound to donor atoms have very long longitudinal relaxation time $T_1$ due to the inversion symmetry of the semiconductor host as well as due to the localized nature of their orbital states. This makes these electron spin states an excellent choice for qubits of a quantum computer. However, the transverse relaxation of the spin polarization in these systems, to the best knowledge of the authors of the work, has never been carefully investigated. 

We calculate the dynamics of the off-diagonal density matrix elements of the electronic spin due to its interaction with acoustic phonons. These density matrix elements represent coherencies of the system intrinsic for a quantum mechanical superposition state, where individual states enter with certain weights/probabilities and phase shifts. Interaction with environment, i.e. phonons, destroys the superposition by randomizing relative phases of the ``up'' and ``down'' states, thus leaving the state of a qubit in a statistical mixture not suitable for quantum computing. 

We find that there are several time scales in the process of such decoherence. The smaller time scale is defined by the size of the orbital localized wave function of the electron (see below). The dephasing of this sort is governed by the interaction with the off-resonant phonons, primarily with wavelength comparable to the effective Bohr's radius of the donor. Decoherence on this time scale - the drop of the off-diagonal elements of the spin density matrix is fast, but incomplete, thus leaving the system in a slightly mixed state. We term this type of decoherence as {\it phase uncertainty} of the state due to the interaction with environment. The later dynamics, i.e., on a much larger time scale, is determined by the conventional relaxation processes associated with spin-flips produced by the resonant phonons studied in [9,10].            

We consider the spin-lattice interaction Hamiltonian for a single electron donor in a semiconductor. Such Hamiltonian can be
generally written as:
$$\eqalign{&{\cal H}_{s-l} = A \mu_B [\sigma_x \tilde \epsilon_{xx} H_x + \sigma_y \tilde \epsilon_{yy} H_y + \sigma_z \tilde \epsilon_{zz} H_z]\cr
&+B \mu_B [\tilde \epsilon_{xy}(\sigma_x H_y + \sigma_y H_x)+ c.p.] + C \mu_B \vec\sigma \cdot{\bf  H} \tilde \Delta\, , \cr} \eqno (1)$$
\noindent
where $\mu_B$ is Bohr's magneton, $\sigma$'s are Pauli matrices, $\epsilon_{\alpha\beta}$ is the effective strain tensor averaged over the ground state effective mass function of the donor, $\tilde \Delta$ is the effective dilatation and {\it cp} stands for "cyclic permutation" [9]. Constants $A$, $B$ and $C$ depend on the specific properties of the host semiconductor as well as of the donor atom. For simplicity we assume that the magnetic field is parallel to the [001] crystal axis, i.e., $H_x=H_y=0$. In this case there is no spin relaxation due to the first term in Eq. (1). Such alignment of the magnetic field can be easily achieved experimentally. The anisotropy of the spin relaxation rate depending on the direction of the magnetic field for P donors in Si has been directly observed [8]. Then the Hamiltonian (1) is reduced to the following form (in interaction representation):
$${\cal H}_{s-l}(t) = \sigma_z Q_z(t) + \sigma_+ Q^-(t) +\sigma_- Q^+(t)\, , \eqno (2)$$
\noindent
where $\sigma_{\pm}=(\sigma_x \pm i\sigma_y)/2$ and 
$$Q_z(t) = \mu_B H [A \tilde \epsilon_{zz}(t) + C \tilde \Delta (t)]\, , \eqno (3)$$
$$Q^{\pm}(t) = B \mu_B H e^{\pm i\omega_s t}[\tilde \epsilon_{zx}(t) \mp i \tilde \epsilon_{yz} (t)]\, , \eqno (4)$$
\noindent
with $\omega_s = g\mu_B H$, where $g$ is the effective electronic $g$-factor of the host semiconductor. It is assumed here that the $g$-factor that determines the Zeeman splitting for the donor electron is the same as that for a free electron. 

The effective strain tensor in (2)-(4) can be expanded in waves with definite wave vectors ${\bf q}$ and polarizations ${\bf n}_p$
$$ \tilde \epsilon_{\alpha\beta} = \sum_{{\bf q}, {\bf n}_p} f({\bf q}, {\bf n}_p) \epsilon_{\alpha\beta}({\bf q},{\bf n}_p)\, , \eqno (5)$$
\noindent
where $\epsilon_{\alpha\beta}({\bf q}, {\bf n}_p)$ can be written in terms of phonon creation/annihilation operators as 
$$\epsilon_{\alpha\beta}({\bf q}, {\bf n}_p) = \left({\hbar \over{8\rho V \omega_{{\bf q}, {\bf n}_p}}}\right)^{1\over2}
\left[( n_{p\alpha} q_{\beta} + n_{p\beta} q_{\alpha})e^{- i \omega_ {{\bf q}, {\bf n}_p} t} a_{{\bf q}, {\bf n}_p}^{\dagger} + H.c. \right]\, .\eqno (6)$$
\noindent
In Eq. (6), $[a_{{\bf q}, {\bf n}_p}, a_{{\bf q^{\prime}}, {\bf n^{\prime}}_p}^{\dagger}] = \delta_{{\bf q},{\bf q^{\prime}}} \delta_{{\bf n}_p , {\bf n^{\prime}}_p}$, $\omega_ {{\bf q}, {\bf n}_p}$ is the phonon frequency, $\rho$ is the crystal mass density, $V$ is the normalization volume, and $\it H.c.$ denotes the Hermitian conjugate part. Function $f({\bf q})$ is the matrix element of the wave with definite ${\bf q}$ taken between the effective mass functions of the donor [10]. Only those lattice vibrations that have wavelength comparable or less than the spatial spread of the donor electron wave function contribute to the interaction (1). In the spherical donor ground state approximation $f({\bf q}) = 1/[1+a_B^2 q^2/4]^2$, where $a_B$ is the effective Bohr radius for the donor [10]. Even though this is not the case for many semiconductors, such as Si or Ge, we will use the above form of $f({\bf q})$ for order of magnitude estimates.

We are interested in the evolution of the reduced density matrix of the electronic spin, $\rho = {\rm Tr}_{ph}\,(\rho_{s+ph})$, with trace taken over the phonon degrees of freedom. Here $\rho_{s+ph}$ is the density matrix of the full system whose evolution is governed by the interaction Hamiltonian (2)-(4). The temporal evolution of the reduced density matrix can be described by the Markovian master equations [11]:
$$\dot \rho (t) = -{1 \over \hbar^2}\sum_{ij}\int_o^t d\tau [(\sigma_i \sigma_j \rho (t) - \sigma_j \rho (t) \sigma_i)
\langle Q_i(t)Q_j(\tau)\rangle $$
$$-(\sigma_i \rho (t) \sigma_j - \rho (t) \sigma_j \sigma_i)\langle Q_j(\tau)Q_i(t)\rangle]\, . \eqno (7)$$
\noindent
In (6) $i,j=z,+,-$, while $\langle \ \rangle$ indicates averages taken over phonon occupation numbers, wave vectors and polarizations. Equations (7) are generally valid on a timescale large compared to the correlation time of the phonon bath.
However, it can be shown that when only the $\sigma_z$ term is retained in the Hamiltonian (2), the decoherence, i.e. the decay of the off-diagonal elements of the spin density matrix, for the exactly solvable model [12] coincides with the solution obtained from (7). Thus we expect that the Markovian assumption built in (7) will still be applicable for the short time regime. Moreover, for the specific case of P donor in Si crystal considered below, the dimensionless constant $A$ is an order of magnitude greater than $B$ and therefore the mechanism of decoherence resulting from coupling to $\sigma_z$ only can be considered separately. The master equation (7) allows us to combine both mechanisms, i.e. phase relaxation via the diagonal and off-diagonal spin components, into a single framework and so in what follows we will pursue with the solution of (7).      
   
In general, equation (7) represents four coupled linear differential equations. Their solution is a cumbersome analytical problem. These equations, however, are greatly simplified if one assumes that the semiconductor host medium is elastically isotropic. Such assumption is commonly used in evaluation of phonon related relaxation processes. For the isotropic phonons the only non vanishing correlators in (7) are $\langle Q_z(t)Q_z(\tau)\rangle$, $\langle Q_-(t)Q_+(\tau)\rangle$ and $\langle Q_+(t)Q_-(\tau)\rangle$. As a result, the equations for the off-diagonal elements decouple from the equations for the diagonal density matrix elements. For the off-diagonal element $\rho_{12}$ one easily obtains that
$$\rho_{12}(t) = \rho_{12}(0)\exp{[-\Gamma (t)]} \eqno (8)$$
with
$$\Gamma (t) = {1 \over \hbar^2} \int_0^t dt'\int_0^{t'} d\tau \left(2\langle [Q_z(t'),Q_z(\tau)]_+\rangle
+\langle [Q_-(t'),Q_+(\tau)]_+\rangle\right)\, .  \eqno (9)$$
\noindent
We evaluate correlators entering (9) for P donors in Si. It was shown by Roth [9] that there are two major mechanisms contributing to the donor electron spin relaxation and thus determining the interaction Hamiltonian (1),(2): $g$-factor modulation due to inter-valley mixing caused by the phonon induced strain as well as $g$-factor modulation in each valley [9]. The first mechanism determines constants $A$ and $C$ in (1),(2). The estimates of [9,10] for Si:P give $A^2 \sim 10$ and $C \simeq -A/3$. The second mechanism is responsible for Hamiltonian with coefficient $B$ in (1), (2).
The estimate in [9] gives $B^2 \simeq 10^{-1}$. While the first term clearly dominates at short times, it saturates for longer times and so the second mechanism prevails. Therefore the first two correlators in (9) that are proportional to $A^2$ are evaluated exactly, while the last two terms (with coefficient $B^2$) are evaluated in secular approximation (with only spin flips contributing) that is valid only for long times [11]. After performing averaging over polarizations and directions of phonon wave vector ${\bf q}$ one obtains that the real part of $\Gamma (t)$ is given by 
$$ {\rm Re}\,\Gamma (t) \simeq \Gamma_0 \gamma_z(t) + {t \over \tau_R}\, ,\eqno (10)$$
where      
$$\Gamma_0 =  {{4 A^2\mu_B^2 H^2 }\over{15 \pi^2 a_B^2 \rho c_t^3\hbar}}\, ,\eqno (11)$$
$$\gamma_z(t)= {a_B^2 \over 2}\int_0^{\infty} dq q f^2(q)\left(1-\cos{(c_ttq)}\right)\coth{\left({\hbar c_t q \over kT}\right)} $$
$$+{a_B^2 c_t^3 \over 3 c_l^3}\int_0^{\infty} dq q f^2(q)
\left(1-\cos{(c_ltq)}\right)\coth\left({{\hbar c_l q }\over{ kT}}\right)\, ,\eqno (12)$$
\noindent
and
$$ \tau_R^{-1} = {{4B^2 (\mu_B H)^5}\over{5\pi\hbar^4\rho}}\left[{1\over{c_t^5}} + {2\over{3 c_l^5}}\right]\coth\left({{\mu_B H}\over{k_BT}}\right)\, . \eqno (13)$$
\noindent
In (10)-(13) $c_l$ and $c_t$ are longitudinal and transverse sound velocities, $T$ is the temperature. For silicon $c_l \ge c_t$ and so the last terms in (12) and (13) can be neglected for an order of magnitude estimate. The integrals (12) converge due the presence of $f(q)$ which imposes a natural cut-off $q \leq a_B^{-1}$ on the magnitude of the wave vector $q$. At the same time in evaluation of $\tau_R^{-1}$ given by (13), $f(q)$ does not contribute because only resonance phonons conditioned by $\hbar c_t q_{res} = 2\mu_B H$ enter in (13), and $f(q_{res}) \simeq 1$. Note that in Eqs. (11) and (13) we also assumed that the average $g$-factor is approximately 2, which is the case for Si.  It also should be emphasized that at low temperature (for quantum computing applications we assume that $T \simeq 10$ mK) $\coth{\left(\hbar c_l q /kT \right)} \simeq 1$ over the range of integration $0 \le q \le a_B^{-1}$. Then it can be seen that the function $\gamma_z(t)$ approaches its maximum value equal {1/3} for $t \ge a_B/c_t \sim 10^{-12}$ s, meaning that the process is very fast and therefore in an applied magnetic field the phase uncertainty of order $\Gamma_0/3$ is practically always present. A similar fast, but incomplete decoherence was found in the study of phase relaxation of nuclear spins in quantum Hall ferromagnet [13]. We argue that this phase uncertainty generated by the first term in (11) is common for most phonon-assisted spin decoherence mechanisms and should be accounted for when evaluating whether a system is a good candidate for quantum computing. 

For $H = 3$ T, the magnetic field typical for quantum computing architectures [3,7], we find for P donor in Si that $\Gamma_0/3 \sim 10^{-9}$ and $\tau_R \sim 1$ s; see Fig. 1.  Note that our Eq. (13) for the conventional spin relaxation has a different from Refs. [9,10] dependence on the applied magnetic field, $H^5$ instead of $H^4$ in [9,10], which comes from the fact that we are interested in a very low, effectively zero temperature regime. Indeed, for high temperatures in (13), $\coth\left({\mu_B H/k_B T}\right) \simeq \left(k_B T /\mu_B H\right)$ and we recover results of [9,10]. Our results indicate that the P donors in Si are excellent candidates for qubits of a quantum computer as both the phase uncertainty and the spin-flip rate satisfy the $10^{-5}$ ``clock speed to decoherence'' criterion [14] for reliable quantum computing. 

Another aspect not considered so far is decoherence due to the interaction of the donor electron spin with optical phonons. Hamiltonian (1) derived in [9,  10] does not include optical phonons as they do not contribute to the real, i.e., spin flip involving relaxation processes, due to the presence of the gap 
$\omega_0$ in their spectrum ($\omega_0$ is optical phonon frequency at $0$ wave vector and $\omega_0 \gg \mu_B H$). Therefore coupling of the optical phonons to $x$ and $y$ components of the spin will create virtually no decoherence. In order to estimate decoherence produced by the optical phonons, we derive the effective coupling of the donor electron spin with optical deformation. The procedure is identical to the derivation of the  Hamiltonian (1) for interaction with acoustic phonons, that was obtained perturbatively in [9,10]. Then we obtain that (see [10] for the details of identical derivation) coupling of $\sigma_z$ to optical phonons is given in terms of the interaction Hamiltonian
$${\cal H}_{opt} = D {\mu_B H \sigma_z} {\langle W_{opt} \rangle \over \Delta} \, , \eqno (14)$$ 
\noindent
where dimensionless constant $D$ is of order $(g_l-g_t)$, the difference between the longitudinal and transverse to the field g-factors, $D \simeq 5 \times 10^{-3}$ for Si, $\langle W_{opt} \rangle$ is the matrix element of the optical deformation potential taken between donor electron effective mass ground state functions and $\Delta$ is the energy splitting between the orbital singlet ground state and the excited doublet states of the donor, $\Delta \sim 10^{-2}$ eV [15]. The decoherence due to the interaction (14) can be evaluated using Eqs. (9)-(12). One obtains that decoherence due to Hamiltonian (14) is also incomplete with $\Gamma_0^{opt} \sim [(g_l-g_t)\mu_B H \Lambda]^2/[\rho \Delta^2 a_B^3 \omega_0^3 \hbar^3]$, where $\Lambda$ is the optical deformation potential constant. For typical parameters of Si one gets that $\Gamma_0^{opt} \sim 10^{-11}$ and so the optical phonons create effectively no decoherence for the localized electron spins in silicon.    
    
One should note that the case of P donor in Si is unique even among the semiconductor crystals with inversion symmetry of the lattice. Indeed the spin-orbit interaction, that mediates interaction of the spin with lattice vibrations, in Ge is several orders of magnitude stronger than in Si. As a consequence the effective coupling constants $A$ and $B$ in the Hamiltonians (1), (2) increase by several orders of magnitude [9,10], resulting in a much higher, presumably by a factor $10^6$, relaxation rate and initial decoherence. Finally, it should be pointed out that the nuclear spins can significantly contribute to decoherence of localized electronic spins by creating fluctuating magnetic fields via the diffusion of the nuclear spin polarization due to nuclear dipole-dipole coupling [16]. Such mechanism, though ineffective for longitudinal spin relaxation ($T_1$) because of large difference in electron and nuclear Zeeman energies, can be important in phase relaxation processes, and should be subjected to detailed study.    

D.M. is thankful to V. Privman and E. Yablonovitch for stimulating discussions. This work  was supported by the Department of Energy (DOE) under contract W-7405-ENG-36, by the National Security Agency (NSA) and Advanced Research and Development Activity (ARDA). One of the authors (D.M.) was supported, in part, by the National Science Foundation grant ECS-0102500.      

\vfil\eject

\centerline{\bf References}

\ 

{\frenchspacing

\item{1.} D.P. DiVincenco, Science {\bf 270}, 255 (1995).

\item{2.} J.M. Kikkawa and D.D. Awschalom, Nature (London) {\bf 397}, 139 (1999); J.M. Kikkawa and D.D. Awschalom, Science {\bf 287}, 473 (2000).

\item{3.} B.E. Kane, Nature {\bf 393}, 133 (1998); B. Koiller, X. Hu, S. Das Sarma, {\it Exchange in silicon based quantum computer architecture}, preprint cond-mat/0106259.

\item{4.} B.E. Kane, N.S. McAlpine, A.S. Dzurak, R.G. Clark, G.J. Milburn, H.B. Sun, H. Wiseman, Phys. Rev. B {\bf 61}, 2961 (2000); A.N. Korotkov Phys. Rev. B {\bf 63}, 115403 (2001); G.P. Berman, F. Borgonovi, G. Chapline, S.A. Gurvitz, P.C. Hammel, D.V. Pelekhov, A. Suter, V.I. Tsifrinovich, {\it Formation and Dynamics of a Schrodinger-Cat State in Continuous Quantum Measurement}, preprint quant-ph/0101035. 

\item{5.} M. Thorwart and P. Hanggi, {\it Decoherence and dissipation during a quantum XOR gate operation}, preprint cond-mat/0104513; A. Garg
Phys. Rev. Lett. {\bf 77}, 764 (1996); D. Mozyrsky, V. Privman, I.D. Vagner, Phys. Rev. B {\bf 63}, 085313 (2001). 

\item{6.} J.M. Kikkawa, I.P. Smorchkova, N. Samarth and D.D. Awschalom, Science {\bf 277}, 1284 (1997); J.M. Kikkawa and D.D. Awschalom, Phys. Rev. Lett. {\bf 80}, 4313 (1998). 
 
\item{7.} D. Loss and D.P. DiVincenzo, Phys. Rev. A {\bf 57}, 120 (1998); V. Privman, I.D. Vagner and G. Kventsel, Phys. Lett. A {\bf 239}, 141 (1998);
A. Imamoglu, D.D. Awschalom, G. Burkard, D.P. DiVincenzo, D. Loss, M. Sherwin and A. Small, Phys. Rev. Lett. {\bf 83}, 4204 (1999); R. Vrijen, E. Yablonovich, K. Wang, H.W. Jiang, A. Balandin, V. Roychowdhury, T. Mor and D.P. DiVincenzo, Phys. Rev. A {\bf 62}, 012306 (2000).

\item{8.} G. Feher and E.A. Gere, Phys. Rev. {\bf 114}, 1245 (1959); A. Honig and E. Stupp, Phys. Rev. Lett. {\bf 1}, 275 (1958), D.K. Wilson and G. Feher, Phys. Rev. {\bf 124}, 1068 (1961).

\item{9.} L.M. Roth, Phys. Rev. {\bf 118}, 1534 (1960); L.M. Roth, Massachusets Institute of Technology Lincoln Laboratory Reports, April 1960 (unpublished).

\item{10.} H. Hasegawa, Phys. Rev. {\bf 118}, 1523 (1960). 

\item{11.} K. Blum, {\it Density Matrix Theory and Applications} (Plenum Press, New York and London, 1996).

\item{12.} G.M. Palma, K.-A. Suominen, A.K. Ekert, Proc. Roy. Soc. London A {\bf 452}, 567 (1996); N.G. van Kampen, J. Stat. Phys. {\bf 78}, 299 (1995);
 D. Mozyrsky and V. Privman, J. Stat. Phys. {\bf 91}, 787 (1998).

\item{13.} T. Maniv, Yu. A. Bychkov, I. D. Vagner, Phys. Rev. B {\bf 64}, 193306 (2001). 

\item{14.} J. Preskill, Proc. Roy. Soc. London A  {\bf 454}, 385 (1998). 

\item{15.} W. Kohn, in {\it Solid-State Physics}, edited by F. Seitz and D. Turnbull (Academic Press, Inc., New York, 1957), Vol. 5, p. 257. 

\item{16.} E. Yablonovitch, private communication. 
    
\vfil\eject
 
\centerline{\bf Figure Caption}

\vskip 0.32in

FIG. 1.\ Schematic dependence of the spin off-diagonal density matrix element on time for P donor in Si. The initial, nonexponential coherence drop on a picosecond scale is followed by exponential, relaxation driven decay.

}

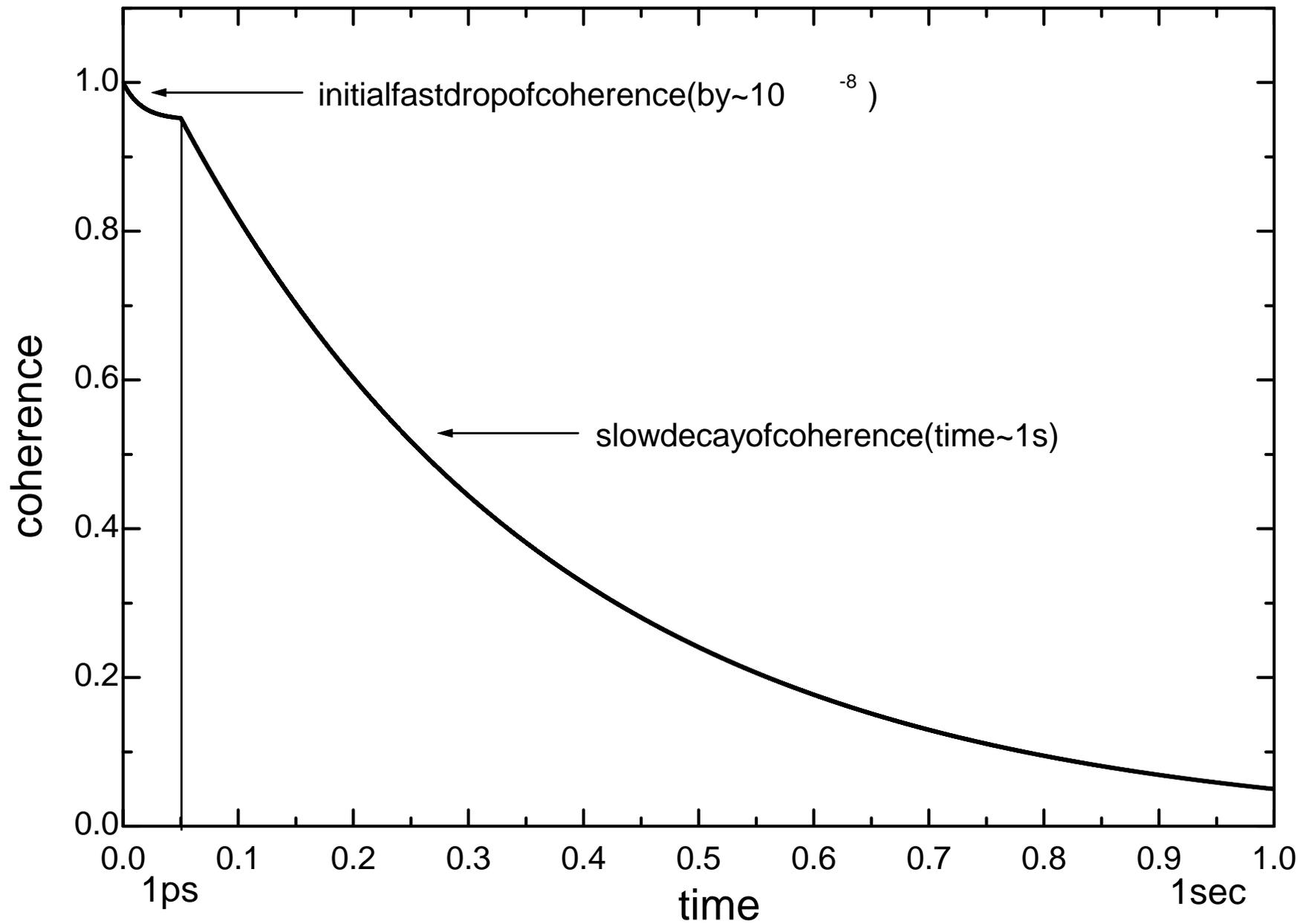
\bye